\documentclass[11pt]{article}
\usepackage{graphicx}
\usepackage[margin=1.1in]{geometry}
\usepackage{bm}
\usepackage{subfig}
\usepackage{natbib}
\usepackage{amsmath,amssymb,amsfonts,amsthm}
\usepackage[noend,ruled,linesnumbered]{algorithm2e}

\newcommand{\N}{\ensuremath{\mathbb{N}}}
\newcommand{\R}{\ensuremath{\mathbb{R}}}

\newcommand{\E}{\ensuremath{\mathbb{E}}}

\newcommand{\Supp}[1]{{Supplementary Information}}

\title{QuickMMCTest -- Quick Multiple Monte Carlo Testing}
\author{Axel Gandy and Georg Hahn
  \\Department of Mathematics, Imperial College London}
\date{}

\begin{document}
\maketitle

\begin{abstract}
Multiple hypothesis testing is widely used to evaluate scientific studies
involving statistical tests.
However, for many of these tests, p-values are not available and are thus
often approximated using Monte Carlo tests such as permutation tests or bootstrap tests.
This article presents a simple algorithm based on Thompson Sampling to
test multiple hypotheses.
It works with arbitrary multiple testing procedures, in particular with step-up and
step-down procedures.
Its main feature is to sequentially allocate Monte Carlo effort,
generating more Monte Carlo samples for tests whose decisions are so far less certain.
A simulation study demonstrates that for a low computational effort,
the new approach yields a higher power and a higher degree of reproducibility of its results
than previously suggested methods.
\end{abstract}

\section{Introduction}
\label{section_thompson_introduction}
Scientific studies are often evaluated by correcting for multiple comparisons.
Several methods published in the literature are used
to correct for multiple tests,
for instance the \cite{Bonferroni1936} correction
or the \cite{Benjamini1995CFD} procedure.

Often, for instance in studies involving biological data, p-values cannot be computed analytically.
They are approximated using Monte Carlo tests such as
permutation tests or bootstrap tests
\citep{Lourenco2014,Martinez2014,Liu2013,Wu2013,Asomaning2012,Dazard2012}.
Permutation tests are widely used in practice as underlying models for biological phenomena
are rarely known.
The evaluation of multiple hypotheses by applying a multiple testing
procedure to Monte Carlo based p-value estimates is the focus of 
this article.

We are interested in evaluating multiple hypotheses using
Monte Carlo samples while ensuring the reproducibility and objectivity of all findings.
\cite{Gleser1996} called this the
\textit{first law of applied statistics}: ``Two individuals using the
same statistical method on the same data should arrive at the same
conclusion.''

We measure this reproducibility in the following way.
We generate a set of fixed p-values for all hypotheses as underlying truth
and consider methods
approximating the p-value of each hypothesis using independent Monte Carlo samples under the null.
Using the approximated p-values,
we aim to reproduce the test result obtained by applying a multiple testing correction
to the fixed p-values.
We are interested in minimising the number of \textit{switched classifications},
that is the number of decisions of individual hypotheses based on estimates which differ
from the ones obtained with the fixed p-values.
Algorithms achieving a low number of switched classifications lead to consistent results
even when applied repeatedly
and thus ensure reproducibility
of their decisions, a feature desired in practice.
Moreover, we aim to achieve a high power in multiple testing in order to obtain meaningful test results,
especially for low computational effort.
The algorithm developed in this article outperforms many existing methods in both regards.

A simple and widely used method to implement a multiplicity correction in the aforementioned scenario
is to draw a constant number of samples per hypothesis,
approximate all p-values
using a conservative p-value estimator
and finally use these estimates as input for the multiple testing procedure,
thus treating them as if they were the p-values
\citep{Nusinow2012,Gusenleitner2012,Rahmatallah2012,Zhou2013,Li2012a}.

The naive approach does not take into account
that hypotheses whose p-values clearly lie in the
rejection or non-rejection area of the multiple testing procedure
should be allocated less samples than
hypotheses whose p-values are closer to the testing threshold and whose decision is thus more difficult to compute.
This leaves considerable scope to improve upon the accuracy of the naive method.

We introduce a sampling algorithm based on Thompson Sampling \citep{Thompson1933}
to compute decisions on multiple hypotheses.
Our approach, called
\texttt{QuickMMCTest}, uses a Beta-binomial model on each
p-value to adaptively decide which hypotheses need to receive
more and which need less samples to obtain fairly clear decisions
(rejections and non-rejections) on all tests.
It avoids computing discrete p-value estimates at any stage,
thus circumventing imprecisions observed in methods using such discrete estimates.
The algorithm works with a variety of commonly used
multiple testing procedures at both constant testing thresholds
as well as variable testing thresholds, that
is thresholds which are functions of the unknown p-values underlying the tests.

The article is organised as follows.
Section \ref{section_thompson} starts by introducing the set-up and presents \texttt{QuickMMCTest}.
In Section \ref{section_example} we discuss a real-data application using gene expression data.
We show that rejections computed with existing methods can lead to high uncertainty concerning
the significance of individual hypotheses.

A simulation study (presented in Section \ref{section_simulation_study})
shows that in comparison to the naive approach,
\texttt{QuickMMCTest}
yields considerably less switched classifications
for popular multiple testing procedures
\citep{Bonferroni1936,Sidak1967,Holm1979,Simes1986,Hochberg1988,
Benjamini1995CFD,BenjaminiYekutieli2001}.

As highlighted in a second simulation study in Section \ref{subsection_simulation_methods_constant_threshold},
the main advantage of the \texttt{QuickMMCTest} algorithm in comparison to existing methods
\citep{BesagClifford1991,GuoPedadda2008,
Sandve2011,JiangSalzman2012,GandyHahn2014}
is its better finite sample behaviour, thus achieving the same accuracy with less computational effort.
However, in contrast to \texttt{MMCTest} of \cite{GandyHahn2014},
it does not provide any guarantees on the correctness of its results.

Section \ref{subsection_simulation_power} conducts power studies.
We show for selected multiple testing procedures
that \texttt{QuickMMCTest} yields a higher power
than the naive method
and than the aforementioned existing methods, especially for low sample sizes.

We conclude with a discussion in Section \ref{section_thompson_discussion}.
Supplementary Material is available for this article which
contains further simulation studies
for a variable testing threshold
as well as an assessment of the dependence of \texttt{QuickMMCTest} on its parameters.
The \texttt{QuickMMCTest} algorithm is implemented in an $R$ package
(\texttt{simctest}, available on CRAN, the Comprehensive R Archive Network).

\section{Methods}
\label{section_thompson}

\subsection{The setting}
\label{subsection_setting}
We would like to test $m$ hypotheses $H_{01}, \ldots, H_{0m}$ for
statistical significance, for each of which we have a statistical test
(and some data) available.
The hypotheses are evaluated using a multiple testing
procedure given by a mapping \citep{GandyHahn2014}
\begin{align*}
h: [0,1]^m \times [0,1] \rightarrow \mathcal{P}(\{ 1,\ldots,m \})
\end{align*}
which takes a vector of $m$ p-values $p \in [0,1]^m$ and a
threshold $\alpha \in [0,1]$ and returns the set of indices of
hypotheses to be rejected, where $\mathcal{P}$ denotes the power set.
Amongst others, the procedures of
\cite{Bonferroni1936}, \cite{Holm1979}, \cite{Shaffer1986}, \cite{Simes1986},
\cite{Hochberg1988}, \cite{Rom1990}, \cite{Benjamini1995CFD} as well as \cite{BenjaminiYekutieli2001}
fit into this framework.

We assume that the p-values $p^\ast=(p_1^\ast,\ldots,p_m^\ast)$ of the
tests for $H_{01}, \ldots, H_{0m}$ are not available analytically.
Instead, we assume that it is possible to draw samples under each null hypothesis.
For each of the samples, we can compute the test statistic and compare
it to the observed test statistic, thus enabling us to approximate the p-values.
We denote the total number of samples drawn for $H_{0i}$ by
$k_i$ and the total number of exceedances of the sampled test
statistic over the observed test statistic among these $k_i$ samples
by $S_i$, where $i \in \{ 1,\ldots,m \}$.
Moreover, the threshold $\alpha$
is allowed to either be constant $\alpha(p^\ast)=\alpha^\ast \in \R$
or a function $\alpha(p^\ast)$ of the p-values $p^\ast$.
In the latter case, $\alpha^\ast$ is itself unknown.

\subsection{The \texttt{QuickMMCTest} algorithm}
\label{subsection_quickmmctest_algorithm}
\texttt{QuickMMCTest}
(Algorithm \ref{algorithm_quickmmctest})
is based on an idea related to
Thompson Sampling \citep{Thompson1933,AgrawalGoyal2012}
and updates a Beta-Binomial model for each p-value in each iteration.

Starting with a Beta$(1,1)$ prior on each p-value,
observing $S_i$ exceedances among $k_i$ samples
results in a Beta$(1+S_i,1+k_i-S_i)$ posterior.
In each iteration of the algorithm, all $m$ p-values are resampled from the posterior distributions,
the multiple testing procedure is evaluated on the $m$ resampled p-values
and the decision of each hypothesis is recorded.
Repeating the above a fixed number of $R$ times allows to compute
an empirical probability that each $H_{0i}$, $i \in \{ 1,\ldots,m \}$,
is rejected ($p^r_i$) and non-rejected ($1-p^r_i$).
The quantity $w_i=\min(p^r_i,1-p^r_i)$ can then be viewed
as a stability measure for the current decision on $H_{0i}$, where $i \in \{ 1,\ldots,m \}$.
We weight rejections and non-rejections equally
when computing the weights $w_i$ in line $10$ of Algorithm \ref{algorithm_quickmmctest}.
However, one might be interested in weighting the rejections $r_i$ and non-rejections $R-r_i$ differently
and incorporate this into the computation of the weights.

\texttt{QuickMMCTest} sequentially draws samples for all hypotheses.
The number of further samples drawn for each $H_{0i}$
is proportional to $w_i$ in each iteration, where $i \in \{ 1,\ldots,m \}$.
This ensures that hypotheses already having a very stable decision
only receive few new samples.

\texttt{QuickMMCTest} has five parameters chosen by the user.
The first two are the total number of samples $K \in \N$ the algorithm
is allowed to spend as well as the algorithm's total number of iterations $n_{\max} \in \N$
(and thus the number of posterior updates).
These two parameters determine $\Delta=K/n_{\max}$, the number of samples allocated in each iteration.
Alternative approaches in which $\Delta$ varies over time are possible.
Furthermore, the parameter $R \in \N$ needs to be provided which controls
the number of replicates used to estimate the weights $w_i$ and
\texttt{QuickMMCTest} depends on the choice of the multiple testing procedure
$h$ as defined in Section \ref{subsection_setting}.
Last, the testing threshold is provided as a function
$\alpha: [0,1]^m \rightarrow \R$ which may either be constant
(that is, $\alpha(p)=\alpha^\ast \in \R$ independent of $p$, where $\alpha^\ast$ is a known constant)
or a function of the p-values $p=(p_1,\ldots,p_m)$.
In the latter case,
the threshold function $\alpha$ is used in Algorithm \ref{algorithm_quickmmctest}
to compute a point estimate of the unknown testing threshold $\alpha(p^\ast)$
in each iteration of the inner loop computing the weights (lines $6-9$).

\begin{algorithm}
\caption{\texttt{QuickMMCTest}}
\label{algorithm_quickmmctest}
\SetKwInOut{Input}{input}
\SetKwFor{Loop}{repeat}{}{end}
\Input{$K$, $n_{\max}$, $R$, $h$, $\alpha$}
$\Delta \leftarrow \lfloor K/n_{\max} \rfloor$,
$k_i \leftarrow 0$, $S_i \leftarrow 0$ for all $i \in \{ 1,\ldots,m \}$\;
\For{$n \leftarrow 1$ \KwTo $n_{\max}$}{
  \lIf{$n=1$}{
    $w_i \leftarrow 1/m$ for all $i \in \{ 1,\ldots,m \}$
  }
  \Else{
    $r_i \leftarrow 0$ for all $i \in \{ 1,\ldots,m \}$\;
    \Loop{$R$ \textnormal{times}}{
      $p_i$ $\sim$ Beta$(1+S_i,1+k_i-S_i)$ independently for all $i \in \{ 1,\ldots,m \}$\;
      For all $i \in \{ 1,\ldots,m \}$: if $i \in h(p,\alpha(p))$ then $r_i \leftarrow r_i + 1$, where $p=(p_1,\ldots,p_m)$\;
    }
    $w_i \leftarrow \min(r_i/R,1-r_i/R)$ for all $i \in \{ 1,\ldots,m \}$\;
    \lIf{$\sum_{j=1}^m w_j=0$}{
      $w_i \leftarrow 1/m$, $i \in \{ 1,\ldots,m \}$
    }
  }
  Use residual sampling with weights proportional to
  $(w_1,\ldots,w_m)$ to decide how to distribute the $\Delta$
  additional samples among the hypotheses\;
  Draw the $\Delta$ samples and update all $k_i$, $S_i$, $i \in \{ 1,\ldots,m \}$\;
}
\Return{$(S_1,\ldots,S_m)$, $(k_1,\ldots,k_m)$}\;
\end{algorithm}

\texttt{QuickMMCTest} uses residual sampling \citep{LiuChen1998}
to guarantee a deterministic minimal allocation of samples to each hypothesis.
After normalising the weights $w_i$,
we first draw $\lfloor w_i\Delta \rfloor$
samples for each $H_{0i}$, $i \in \{ 1,\ldots,m \}$.
The remaining $\Delta - \sum_{j=1}^m \lfloor w_j\Delta \rfloor$
samples are then allocated one sample at a time with weights proportional to
$(w_1\Delta - \lfloor w_1\Delta \rfloor, \ldots, w_m\Delta - \lfloor w_m\Delta \rfloor)$.

Alternatively, one could replace the residual sampling
by simple multinomial sampling or other methods used in,
for instance, particle filters.

Calculating the weights is computationally fast
as it only requires $R$ draws from each of the $m$ Beta distributions as opposed to drawing samples
from the data (for instance via permutations which can be costly).

Decisions on all hypotheses can be obtained in various ways with \texttt{QuickMMCTest}.
Naively, one could compute $h( \hat{p},\alpha(\hat{p}) )$,
where $\hat{p}=(\hat{p}_1,\ldots,\hat{p}_m)$ is a vector of estimates
$\hat{p}_i = (S_i+1) / (k_i+1)$, $i \in \{ 1,\ldots,m \}$,
computed with a pseudo-count \citep{DavisonHinkley1997}.

We do not consider unbiased p-value estimates $S_i / k_i$, $i \in \{ 1,\ldots,m \}$,
computed without a pseudo-count as
such estimates lead to tests not keeping the prescribed error level
\citep{DavisonHinkley1997,Manly1997,EdgingtonOnghena2007}.

A more sophisticated approach to obtain final rejections and non-rejections
is to recompute decisions on all hypotheses $R$ times using
draws from the final Beta posteriors after termination of Algorithm \ref{algorithm_quickmmctest}.
Recording the number of rejections $r_i$ per hypothesis
allows to compute empirical rejection probabilities
as done for the computation of the weights $w_i$
in lines $6$ to $9$ of Algorithm \ref{algorithm_quickmmctest}, where $i \in \{ 1,\ldots,m \}$.
Each hypothesis $H_{0i}$, $i \in \{ 1,\ldots,m \}$, is rejected
if and only if $r_i/R>0.5$,
that is if $H_{0i}$ was predominantly rejected based on resampled p-values.
The cutoff of $0.5$ is arbitrary and can be replaced by higher (lower) values to make
\texttt{QuickMMCTest} more (less) conservative.

We recommend computing decisions for all hypotheses using the latter approach based on empirical rejection
probabilities as such test results contain less switched classifications
and hence ensure a higher degree of reproducibility
than the ones based on p-value estimates.
We demonstrate this in the Supplementary Material.
Moreover, \texttt{QuickMMCTest} with empirical rejection probabilities
has a higher power than its variant with point estimates (see Supplementary Material).

In the simulation studies of
Section \ref{section_simulation_study}
we employ \texttt{QuickMMCTest} with parameters $n_{\max}=10$ and $R=1000$
and determine decisions on all hypotheses using empirical rejection probabilities computed with a cutoff of $0.5$.
In the
Supplementary Material we investigate the choice of these parameters,
showing that there is no strong case to increase $n_{\max}$ and $R$
further as it does not considerably improve performance.

The choice of $K$ and $n_{\max}$ affects the performance of Algorithm \ref{algorithm_quickmmctest}:
To be precise, $n_{\max}$ needs to be large enough (around $n_{\max}=10$ to $n_{\max}=100$)
to allow \texttt{QuickMMCTest} to iteratively adjust the weights
according to the stability of the decision on each hypothesis.
At the same time, $\Delta=K/n_{\max}$ has to be large enough to ensure that
in line $13$ of Algorithm \ref{algorithm_quickmmctest},
new samples are drawn even for hypotheses with very low weights --
this is needed to ensure that especially in the first iterations of Algorithm \ref{algorithm_quickmmctest},
no hypothesis is excluded preliminarily from receiving new samples in line $13$.
The influence of $K$ and $n_{\max}$ (and thus of $\Delta=K/n_{\max}$)
on the accuracy of \texttt{QuickMMCTest}
is exemplarily demonstrated in the Supplementary Material.

For an example run on generated p-values,
the Supplementary Material visualises
how the sample allocation computed by \texttt{QuickMMCTest} compares
to the p-values and the testing threshold.

\section{An application of multiple testing to gene expression data}
\label{section_example}

\begin{figure}
\centering\includegraphics[width=0.5\textwidth]{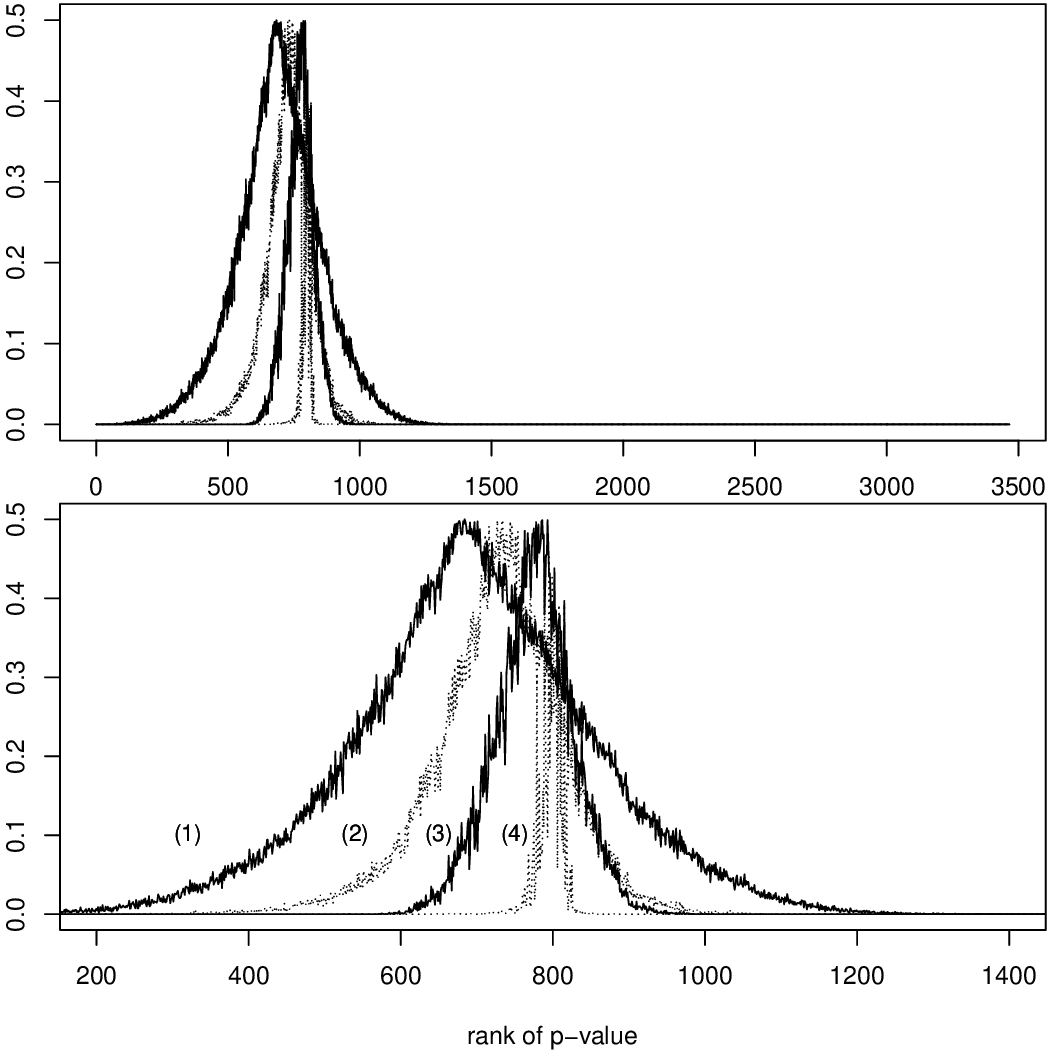}
\caption{Plot of the probability of a random decision $\min (p_i^r,1-p_i^r)$
for each hypothesis ordered by the rank of their p-value
(naive method with $1000$ permutations $(1)$
and $10000$ permutations $(3)$ per hypothesis,
\texttt{QuickMMCTest} with total effort $K=1000m$ $(2)$ and $K=10000m$ $(4)$).
Top shows all hypotheses,
bottom shows a zoomed-in region around the last significant hypothesis.}
\label{fig:intro}
\end{figure}

We consider a dataset of
gene modifications (so-called \textit{H3K4me2}-modifications)
of \cite{Pekowska2010}:
This dataset defines certain regions on a genome and
specifies the midpoints of gene modifications within each region.

We are interested in testing whether the gene modifications
appear more often in the lower half of the gene regions using the test of \cite{Sandve2011}.
For this,
\cite{Sandve2011} first norm the beginning and the end of each
region to $0$ and $1$, respectively.
The midpoints of $k \in \N$ gene modifications then correspond
to $k$ random points $X_1,\ldots,X_k$ in the interval $[0,1]$.
We test the null hypothesis
$H_0: \E(T) \geq 0.5$ against the alternative $H_1: \E(T) < 0.5$
using the test statistic $T=\frac{1}{k} \sum_{i=1}^k X_i$
in connection with a permutation test.
For each gene region and its set of midpoints,
a permutation is generated by permuting the locations of the midpoints in $[0,1]$
while preserving all inter-point distances.
\cite{Sandve2011} first filter the dataset for regions with at least $10$ midpoints
and define each such region to be one hypothesis.
This leads to $m=3465$ hypotheses under consideration.

We are interested in the probability of a random decision,
meaning the probability that a single hypothesis switches between
being significant and non-significant in repeated
test results. For this, we generated
both $1000$ (``low effort``) and $10000$ (``high effort``) permutations per hypothesis,
approximated its p-value with a pseudo-count in both the numerator and denominator
and then tested all hypotheses
by applying the \cite{Benjamini1995CFD} procedure
with threshold $0.1$.
This was repeated $r=2000$ times.

When applying the \cite{Benjamini1995CFD} procedure with a fixed threshold of $0.1$,
the allocation strategy based on Beta posteriors in \texttt{QuickMMCTest}
is not yet used in the first iteration;
instead, an equal weight is placed on each hypothesis (line $3$ of Algorithm \ref{algorithm_quickmmctest}).
Only after the inital batch of $\Delta$ samples is drawn,
the \cite{Benjamini1995CFD} procedure can be applied in all subsequent iterations
to resampled p-values from the Beta posteriors and the constant threshold of $0.1$ (line $8$)
in order to compute new weights (line $10$) and to allocate further samples
(lines $13$ and $14$ of Algorithm \ref{algorithm_quickmmctest}).

As done in Section \ref{subsection_quickmmctest_algorithm}
we quantify the uncertainty in these $r$ test results
by computing empirical probabilities
$p_i^r$ ($1-p_i^r$)
that hypothesis $i$ is rejected (non-rejected) among the $r$ repetitions.
We define $\min (p_i^r,1-p_i^r) \in [0,0.5]$
as the probability of a random decision.

We repeat this experiment with \texttt{QuickMMCTest}
on the same dataset an equal number of times
and compute probabilities of a random decision.
To ensure a fair comparison, we allow at most the same total number of samples the naive method had used, thus
$K=1000m$ for low effort and $K=10000m$ for high effort,
where $m$ is the number of hypotheses.

Figure \ref{fig:intro} displays the probabilities $\min (p_i^r,1-p_i^r)$
of a random decision
for the entire range of genes (top) in the
\cite{Pekowska2010}
dataset as well as for a zoomed-in region around the last significant hypothesis (bottom) occuring within the ranks $600-800$.
These curves correspond to the following methods:
the naive method with $1000$ permutations $(1)$
and $10000$ permutations $(3)$ per gene as well as the
\texttt{QuickMMCTest} algorithm
with a total effort $K=1000m$ $(2)$ and $K=10000m$ $(4)$.
Hypotheses are ordered according to the rank of their p-value estimate computed with $10^7$ permutations.
Due to the finite computational effort and the very low p-values
the ordering of the hypotheses exhibits a certain noise.

Figure \ref{fig:intro} (bottom) shows that \texttt{QuickMMCTest}
yields test results
at a low effort ($K=1000m$) with a level of uncertainty which is comparable to the one of
the naive method at a high effort
($10000$ permutations per hypothesis).
Using $K=10000m$ samples, \texttt{QuickMMCTest} yields test results which are
considerably more stable and contain less random decisions
than the ones of the naive method over the entire range of hypotheses.

Notably, comparing the location of the last rejected hypothesis (and thus the location of the peaks)
in Figure \ref{fig:intro} shows that,
using the same computational effort,
\texttt{QuickMMCTest} is able to reject more hypotheses than the naive method.
This is a desired feature for practical applications.
For datasets with few significant hypotheses, the number of rejections increases
when spending more samples (or when allocating samples more efficiently as in the case of \texttt{QuickMMCTest})
due to the fact that higher numbers of samples
(in connection with a pseudo-count in the numerator and denominator)
give a higher resolution and capture more low p-values below the threshold.

\section{Simulation study}
\label{section_simulation_study}
We evaluate \texttt{QuickMMCTest} on a simulated dataset in three ways.
First, the performance of \texttt{QuickMMCTest} is compared
to the one of the naive method
using a variety of commonly used multiple testing procedures
at a constant testing threshold
(Section \ref{subsection_simulation_study_naive}).
Second, we fix the procedures of \cite{Bonferroni1936} and \cite{Benjamini1995CFD}
as multiple testing procedures and compare \texttt{QuickMMCTest}
to a variety of common methods published in the literature
(Section \ref{subsection_simulation_methods_constant_threshold}).
Third, we conduct power studies for \texttt{QuickMMCTest} and the naive
method for selected multiple testing procedures, showing that \texttt{QuickMMCTest}
yields a higher power even for low samples sizes.

\subsection{The simulation setting}
\label{subsection_simulation_setting}
In order to be able to compute numbers of switched classifications
we need to simulate exceedances of the
sampled test statistic over the observed test statistic. For this it
suffices to fix a set of p-values and simulate exceedances for
the tests by sampling independent Bernoulli random variables with
success probability being equal to the p-value of each test.

In Sections \ref{subsection_simulation_study_naive} and \ref{subsection_simulation_methods_constant_threshold}
we use one fixed set of $m=5000$ p-values.
These p-values are independent realisations from a mixture distribution with a
proportion $0.9$ coming from a Uniform$[0,1]$ distribution and the
remaining proportion $0.1$ coming from a Beta$(0.25,25)$ distribution.
A large proportion of p-values coming from the null would also be expected in practice.
This model was used in \cite{Sandve2011}.

Comparing the test result returned by any algorithm to the result
obtained by applying the multiple testing procedure to the fixed set
of $m$ p-values allows to compute numbers of switched classifications (see
Section \ref{subsection_setting}) with respect to the fixed p-values.

In Section \ref{subsection_simulation_power}, in each repetition of the experiment,
we draw $m$ Bernoulli random variables with probability $0.1$.
These random variables serve as indicators for the falseness of the null hypothesis.
We then draw the p-value for each false null hypothesis from a Beta$(0.25,25)$ distribution,
and all remaining p-values from a uniform distribution in $[0,1]$.
Comparing the decisions on all hypotheses computed by any algorithm
to the falseness indicators thus allows
to compute averages of type I error and power.
In our multiple testing setting, we compute the \textit{per-pair power}, defined as the average
probability of rejecting a false null hypothesis.

All results are based on $1000$ repetitions.
The error of the simulations is less than the least significant digit we report in the tables
presented in this section.

\subsection{Comparison to a naive method for various multiple testing procedures}
\label{subsection_simulation_study_naive}

\begin{table*}
\caption{Average numbers of switched classifications
(average numbers of switched rejections in brackets)
for the naive method compared to \texttt{QuickMMCTest} (Alg.\ \ref{algorithm_quickmmctest})
for common multiple testing procedures. Constant testing threshold $0.1$.}
\label{Tab:comparison_naive_fixed_threshold}
\begin{center}
\begin{tabular}{r||rr|rr}
&\multicolumn{2}{c|}{low effort ($s=1000$)}
&\multicolumn{2}{c}{high effort ($s=10000$)}\\
& naive & Alg.\ \ref{algorithm_quickmmctest}
& naive & Alg.\ \ref{algorithm_quickmmctest}\\
\hline
\cite{Bonferroni1936}   &87 (0) &43.8 (2.6)     &87 (0) &3 (1.7)\\ 
\cite{Simes1986}        &32 (9.6)       &2 (0.9)        &9 (3.8)        &0.1 (0.1)\\ 
\cite{Hochberg1988}     &87 (0) &43.4 (2.5)     &87 (0) &3.2 (2)\\ 
\cite{Benjamini1995CFD} &31.9 (9.5)     &2 (1)  &9.1 (3.8)      &0.1 (0.1)\\ 
\cite{BenjaminiYekutieli2001}   &162 (0)        &14.5 (3.3)     &22 (5.5)       &0.6 (0.6)\\ 
\cite{Sidak1967}        &90 (0) &36.3 (2.9)     &90 (0) &3.5 (1.6)\\ 
\cite{Holm1979} &88 (0) &39.5 (3.2)     &88 (0) &3.4 (2.1)\\
\end{tabular}
\end{center}
\end{table*}

We compare \texttt{QuickMMCTest}
to the naive method (Section \ref{section_thompson_introduction})
for a variety of commonly used multiple testing procedures using the constant threshold $\alpha^\ast=0.1$.
These procedures are the step-up procedures of
\cite{Bonferroni1936}, \cite{Simes1986}, \cite{Hochberg1988},
\cite{Benjamini1995CFD} and \cite{BenjaminiYekutieli2001}
as well as the step-down procedures of
\cite{Sidak1967} and \cite{Holm1979}.

The naive method draws a fixed number of $s$ samples per hypothesis,
estimates each $p_i^\ast$ as $\hat{p}_i=(e_i+1)/(s+1)$
\citep{DavisonHinkley1997}
and returns $h( \hat{p},\alpha(\hat{p}) )$,
where
$e_i$ is the number of exceedances observed for $H_{0i}$, $i \in \{ 1,\ldots,m \}$, among $s$ samples
and $\hat{p} = (\hat{p}_1,\ldots,\hat{p}_m)$.

We repeatedly apply the naive method at both a low effort
(defined as using $s=1000$ samples to estimate the p-value of each hypothesis)
and a high effort ($s=10000$) and in both cases apply
\texttt{QuickMMCTest} with a matched effort.

As stated in Section \ref{subsection_quickmmctest_algorithm},
computing final decisions on all hypotheses with \texttt{QuickMMCTest} by using
empirical rejection probabilities as opposed to point estimates
leads to both less switched classifications as well as a higher power
(shown in the Supplementary Material).
Results in this section
are therefore given for empirical rejection probabilities only.
Nevertheless, analogous results to the ones in Table \ref{Tab:comparison_naive_fixed_threshold}
obtained with \texttt{QuickMMCTest} in connection with point estimates
can be found in the Supplementary Material.

Table \ref{Tab:comparison_naive_fixed_threshold} presents simulation results.
When applying the naive method to the procedures of
\cite{Bonferroni1936} and \cite{Hochberg1988} as well as \cite{BenjaminiYekutieli2001}, \cite{Sidak1967} and
\cite{Holm1979}, the following phenomenon can be observed.
Using a pseudo-count causes all p-value estimates $\hat{p}_i$
to be bounded below by $1/(k_i+1)$, where $i \in \{ 1,\ldots,m \}$.
Due to the low threshold used by the \cite{Bonferroni1936} correction,
this lower bound is larger than the testing threshold itself,
leading to all hypotheses being consistently non-rejected and thus to meaningless results
(see the Supplementary Material for more details).
The number of recorded switched classifications is hence equal to the number of undetected rejections.
For the \cite{BenjaminiYekutieli2001} procedure, results are meaningful only at a high effort.

The naive method is able to compute meaningful results at a low effort
for the two procedures of
\cite{Simes1986} and \cite{Benjamini1995CFD} only,
even though results still contain around $30$ switched classifications on average.
Most importantly, the naive method erroneously rejects considerably more hypotheses than
\texttt{QuickMMCTest} (in the cases where rejections can be observed),
thus reporting more false findings which is undesired in practice.

In contrast to the naive method,
\texttt{QuickMMCTest} does not rely on computing p-value estimates
and therefore computes meaningful results for all procedures at both a low and a high effort.
At a low effort, these are very accurate for the procedures of
\cite{Simes1986} and \cite{Benjamini1995CFD}.
For all other methods, our approach yields around $35$ to $45$ switched classifications.

Applying the naive method at a high effort with $s=10000$ samples per hypothesis
is still not sufficient to observe any rejections for the procedures of
\cite{Bonferroni1936}, \cite{Hochberg1988}, \cite{Sidak1967} or \cite{Holm1979}.
For all other procedures, the naive method yields around $10$ to $20$ switched classifications.

At a high effort,
\texttt{QuickMMCTest} yields considerably less switched classifications
than the naive method for all multiple testing procedures
under consideration and essentially no switched classifications for the two procedures of
\cite{Simes1986} and \cite{Benjamini1995CFD}.
Similarly to the comparison at a low effort,
our algorithm erroneously rejects considerably
less hypotheses than the naive method, a feature desired for practical use.

The results of Table \ref{Tab:comparison_naive_fixed_threshold} are confirmed by a second study using
a variable testing threshold which depends on the unknown p-values.
To be precise, we correct the testing threshold using
$\alpha(p^\ast) = \alpha^\ast/\hat{\pi}_0(p^\ast)$, where $\alpha^\ast=0.1$ and
$\hat{\pi}_0(p)=\min \left( 1, 2/m \sum_{i=1}^m p_i \right)$
is a robust estimate of the proportion $\pi_0$ of true null hypotheses of \cite{PoundsCheng2006}.
The Supplementary Material
shows that the results for this variable testing threshold
are qualitatively similar to the ones in Table \ref{Tab:comparison_naive_fixed_threshold}.

\subsection{Comparison to a variety of common methods}
\label{subsection_simulation_methods_constant_threshold}
We now compare \texttt{QuickMMCTest} to previously suggested
algorithms to test multiple hypotheses based on Monte Carlo sampling.
These algorithms are the naive method and the algorithms of
\cite{BesagClifford1991}, \cite{GuoPedadda2008},
\cite{Sandve2011}, \cite{JiangSalzman2012} as well as \cite{GandyHahn2014}.
All methods are run with standard parameters suggested by their authors
(the precise parameters are also listed in the Supplementary Material).

The  \cite{Bonferroni1936} correction
applied at a constant threshold $\alpha^\ast=0.1$
is used to evaluate the $m=5000$ p-values fixed in Section \ref{subsection_simulation_setting}.
All methods are run at a low and a high effort, where the
naive method is used as a reference to set the total effort $K$.
We define low effort as $K=1000m$, the effort equivalent to
spending $s=1000$ samples per hypothesis, and similarly high effort as $K=10000m$.
Results are displayed in Table \ref{Tab:comparison_methods_fixed_threshold}.

\begin{table}
\caption{Average numbers of switched classifications
(average numbers of switched rejections in brackets)
for common methods compared to \texttt{QuickMMCTest}
using the \cite{Bonferroni1936} correction. Constant threshold $0.1$.}
\label{Tab:comparison_methods_fixed_threshold}
\begin{center}
\begin{tabular}{r||r|r}
& low effort & high effort\\
\hline
Naive method    &87 (0) &87 (0)\\
\cite{BesagClifford1991}        &87 (0) &4.9 (2.4)\\
\cite{GuoPedadda2008}   &87 (0) &4.5 (2.1)\\
\cite{Sandve2011}       &87 (0) &19 (1.6)\\
\cite{JiangSalzman2012} &87 (0) &16.3 (3.7)\\
\cite{GandyHahn2014}    &87 (0) &5 (2.2)\\
\texttt{QuickMMCTest} 	&43.7 (2.6)     &3 (1.7)\\
\end{tabular}
\end{center}
\end{table}

\begin{table}
\caption{Average numbers of switched classifications
(average numbers of switched rejections in brackets)
for common methods compared to \texttt{QuickMMCTest}
using the \cite{Benjamini1995CFD} procedure. Constant threshold $0.1$.}
\label{Tab:comparison_methods_fixed_threshold_BH}
\begin{center}
\begin{tabular}{r||r|r}
& low effort & high effort\\
\hline
Naive method    &31.9 (9.7)     &9.1 (3.7)\\
\cite{BesagClifford1991}        &18.3 (7.5)     &18.3 (7.4)\\
\cite{GuoPedadda2008}   &4.5 (2.1)      &0.3 (0.3)\\
\cite{Sandve2011}       &10 (4) &2.8 (1.3)\\
\cite{JiangSalzman2012} &13.2 (4.9)     &3.5 (1.6)\\
\cite{GandyHahn2014}    &9.9 (4.1)      &0.8 (0.6)\\
\texttt{QuickMMCTest}	&2 (1)  &0.1 (0.1)\\
\end{tabular}
\end{center}
\end{table}

For a low effort,
Table \ref{Tab:comparison_methods_fixed_threshold} demonstrates
that due to the very low threshold of the \cite{Bonferroni1936} correction,
all methods except for \texttt{QuickMMCTest} are unable to compute p-value
estimates with a resolution sufficient to detect any rejections.
They are thus unable to compute meaningful decisions,
leading to switched classification numbers equal to the
$87$ rejections observed when applying the \cite{Bonferroni1936} correction to the fixed p-values.

\texttt{QuickMMCTest} does not suffer from this problem
and yields roughly $40$ (out of $m=5000$) switched classifications with
a very low number of false findings.
If the computation of weights in \texttt{QuickMMCTest} was replaced by alternative approaches
relying on discrete p-value estimates,
\texttt{QuickMMCTest} would be susceptible again to not being able to record any rejections
like the other methods considered in Table \ref{Tab:comparison_methods_fixed_threshold}.

At a high effort, most methods compute acceptable test results
with around $5$ switched classifications with the exception of the naive method
which is still unable to detect any rejections.
\texttt{QuickMMCTest} yields a slightly lower average of switched classifications than the other methods.

Table \ref{Tab:comparison_methods_fixed_threshold_BH} repeats the previous comparison
using the \cite{Benjamini1995CFD} procedure controlling the false discovery rate.
Due to the less conservative nature of the \cite{Benjamini1995CFD} procedure,
all methods are able to compute meaningful test results at both a low and a high effort.
The naive method and the one of \cite{BesagClifford1991} perform poorly in this new scenario.
The method of \cite{GuoPedadda2008} performs very well and is only outperformed by
\texttt{QuickMMCTest} at a low effort
(yielding half as many switched classifications as \cite{GuoPedadda2008} and a multiple fold decrease compared to all other methods).
At a high effort, \cite{GuoPedadda2008} perform comparably to \texttt{QuickMMCTest}.

These results are again consistent for a variable testing threshold
as demonstrated in the Supplementary Material
(using the threshold of \cite{PoundsCheng2006}, see Section \ref{subsection_simulation_study_naive}).

The similar performance of the algorithms of \cite{GuoPedadda2008},
\cite{GandyHahn2014} as well as \texttt{QuickMMCTest}
is not a coincidence.
Both \cite{GuoPedadda2008} as well as \cite{GandyHahn2014} use
a monotonicity property of step-up and step-down procedures
of \cite{TamhaneLiu2008} to stop the sampling process for certain hypotheses
in order to allocate the remaining samples equally to hypotheses
whose decisions are computationally more demanding to compute.
Neither of them uses any weights to fine-tune this equal allocation.

\texttt{QuickMMCTest} is able to both concentrate available
samples on hypotheses whose decisions are harder to compute
as well as to fine-tune this allocation to individual hypotheses using weights,
a feature which yields another improvement
in accuracy compared to the other two methods.

\subsection{\texttt{QuickMMCTest} yields a higher power}
\label{subsection_simulation_power}

\begin{figure*}
\centering
\begin{minipage}{0.4\textwidth}
\includegraphics[width=\textwidth]{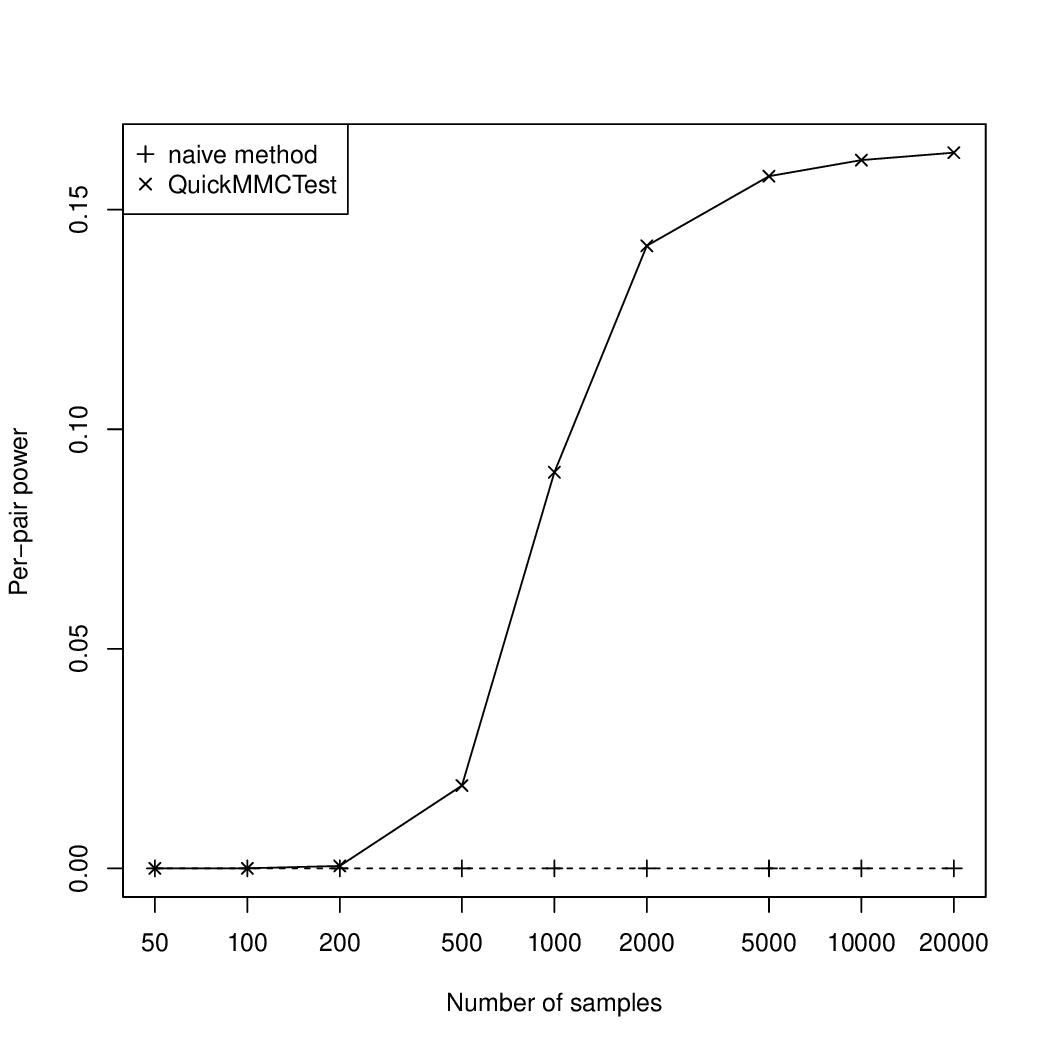}
\end{minipage}~
\begin{minipage}{0.4\textwidth}
\includegraphics[width=\textwidth]{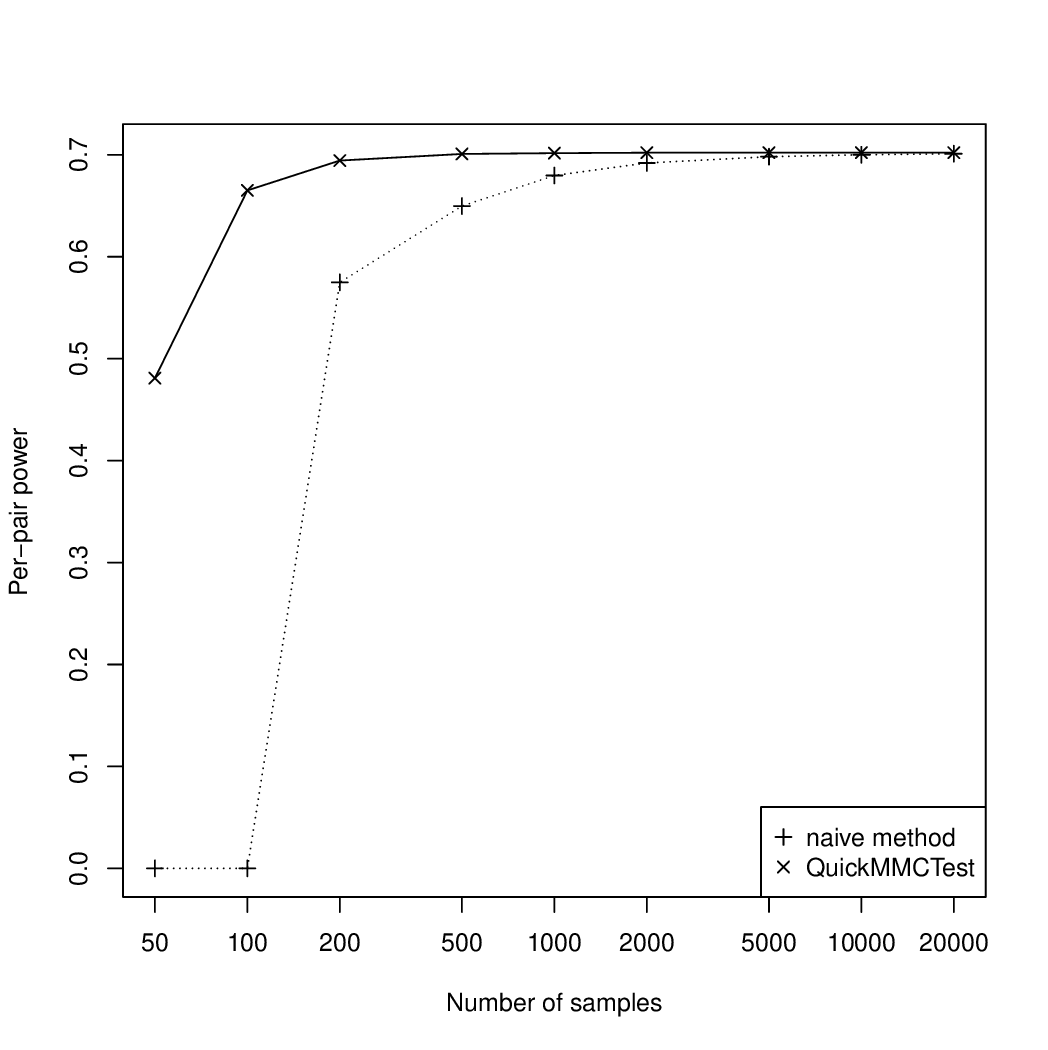}
\end{minipage}
\caption{Average per-pair power against number of samples per hypothesis.
Comparison of the naive method against \texttt{QuickMMCTest} for
the \cite{Bonferroni1936} correction (left) and the \cite{Simes1986} procedure (right).
Log-scale on the x-axis.}
\label{fig:power_B_SS}
\end{figure*}

\begin{figure*}
\centering
\begin{minipage}{0.4\textwidth}
\includegraphics[width=\textwidth]{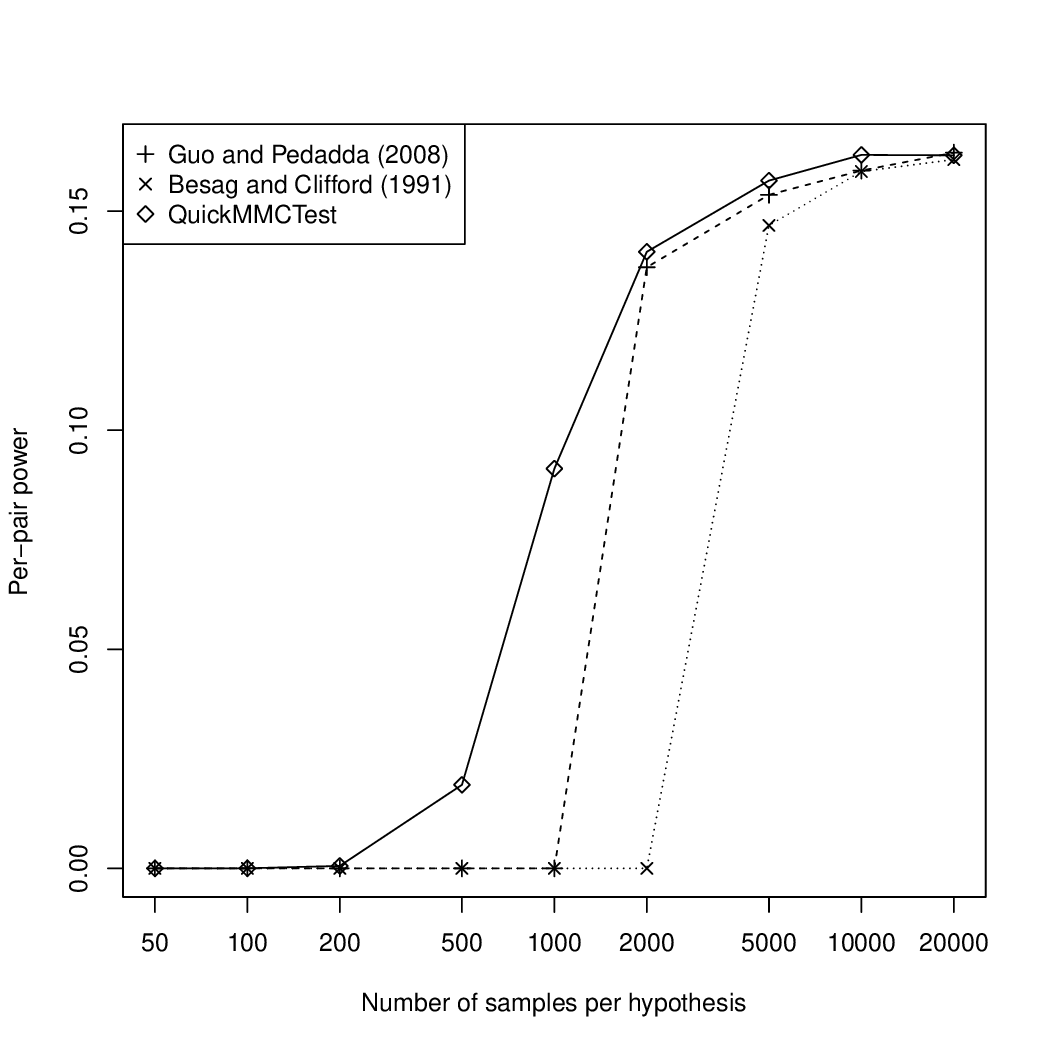}
\end{minipage}~
\begin{minipage}{0.4\textwidth}
\includegraphics[width=\textwidth]{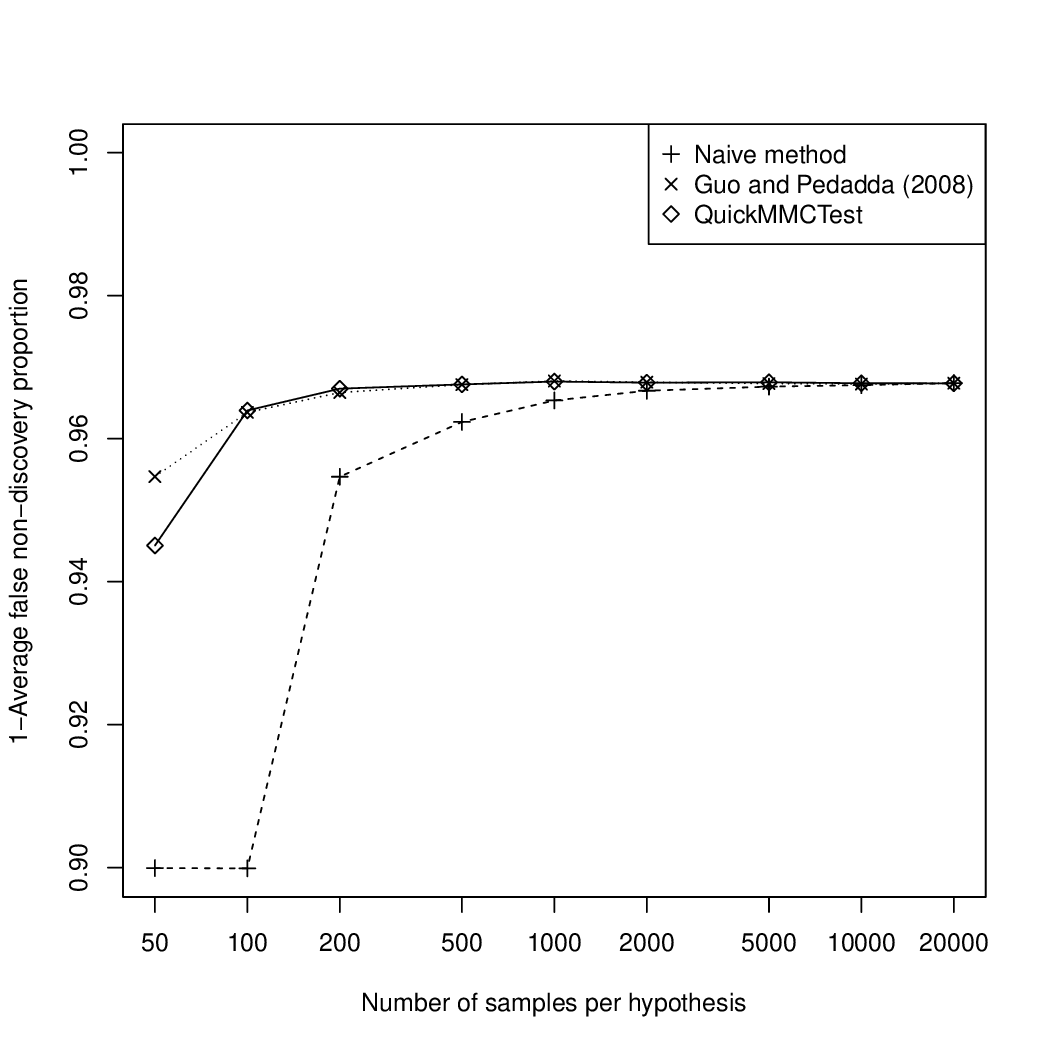}
\end{minipage}
\caption{Left: Average per-pair power against number of samples per hypothesis.
Comparison of the algorithms of \cite{GuoPedadda2008} and \cite{BesagClifford1991} against \texttt{QuickMMCTest}.
Multiple testing with the \cite{Bonferroni1936} correction.
Right: One minus the average false non-discovery proportion against number of samples per hypothesis.
Comparison of the naive method and the algorithm of \cite{GuoPedadda2008} against \texttt{QuickMMCTest}.
Multiple testing with the \cite{Benjamini1995CFD} procedure.
Log-scale on the x-axis.}
\label{fig:power_methods}
\end{figure*}

We compare the power of \texttt{QuickMMCTest} to the one of the naive method and
selected algorithms used in Section \ref{subsection_simulation_methods_constant_threshold}
as a function of the number of samples per hypothesis.
For this we use the simulation setting described in Section \ref{subsection_simulation_setting}.
As in Section \ref{subsection_simulation_study_naive},
\texttt{QuickMMCTest} is applied with a matched total effort.

Figure \ref{fig:power_B_SS} shows the average (per-pair) power of both the naive method
and \texttt{QuickMMCTest} as a function of the number of samples.
As seen previously in Table \ref{Tab:comparison_methods_fixed_threshold},
due to the very low threshold of the \cite{Bonferroni1936} correction,
the naive method is not able to detect any rejections even for
large numbers of samples
(left plot in Figure \ref{fig:power_B_SS}).
Its power is therefore zero.
\texttt{QuickMMCTest} initally suffers from the same problem, but is able to gain power
when using an effort equivalent to $500$ samples per hypothesis onwards.
For the less conservative \cite{Simes1986} procedure
(right plot in Figure \ref{fig:power_B_SS}),
the naive method gains power from $200$ samples per hypothesis onwards.
\texttt{QuickMMCTest} achieves the same power as the naive method
a lot faster with less samples: for instance, the power of \texttt{QuickMMCTest} with $100$
samples per hypothesis is comparable to the one of the naive method with $1000$ samples.

Figure \ref{fig:power_methods}
repeats the power study for two fixed multiple testing procedures
and compares \texttt{QuickMMCTest} to four selected existing methods
already considered in
Section \ref{subsection_simulation_methods_constant_threshold}.

Figure \ref{fig:power_methods} (left) shows that for the \cite{Bonferroni1936} correction,
\texttt{QuickMMCTest} yields a much earlier incease in power for low samples sizes
than the algorithms of \cite{BesagClifford1991} and \cite{GuoPedadda2008},
which
in Table \ref{Tab:comparison_methods_fixed_threshold}
both performed comparably well to \texttt{QuickMMCTest}.

Figure \ref{fig:power_methods} (right) repeats this comparison for the fdr analogue of the
power, precisely $1-$\textit{fnp}, where \textit{fnp} is the false non-discovery rate,
defined as the proportion of false negatives among the accepted null hypotheses.
The plot shows that when controlling the false discoveries using the \cite{Benjamini1995CFD} procedure,
both the algorithm of \cite{GuoPedadda2008} as well as \texttt{QuickMMCTest}
achieve a higher power than the naive approach for low samples sizes,
with a slight advantage for \cite{GuoPedadda2008}.

In all comparisons presented in this section, the procedures kept the
familywise error rate or the false discovery proportion at the $\alpha=0.1$ level, respectively.

The plots for power comparison of the naive method to \texttt{QuickMMCTest}
using the other multiple testing procedures considered in Section \ref{subsection_simulation_study_naive}
are qualitatively similar to the ones in Figure \ref{fig:power_B_SS}.
Similar to the left plot in Figure \ref{fig:power_methods},
\texttt{QuickMMCTest} also outperforms all other methods considered
in Section \ref{subsection_simulation_methods_constant_threshold}
in terms of the per-pair power
when controlling the familywise error.
With the exception of \cite{GuoPedadda2008},
the same holds true when comparing \texttt{QuickMMCTest} to the methods considered
in Section \ref{subsection_simulation_methods_constant_threshold} in terms of $1-$\textit{fnp}
similar to the right plot in Figure \ref{fig:power_methods}.

All comparisons in this article
use \texttt{QuickMMCTest} with empirical rejection probabilities to determine decisions.
The Supplementary Material repeats
the power studies of Figures \ref{fig:power_B_SS} and \ref{fig:power_methods}
when employing \texttt{QuickMMCTest} with both empirical rejection probabilities as well as point estimates
to obtain decisions on all hypotheses after stopping.
We show that empirical rejection probabilities lead to a higher power, especially for low computational effort.

\section{Discussion}
\label{section_thompson_discussion}
We considered multiple testing in a realistic scenario in which
it is not possible to compute p-values analytically for all tests.
Instead, we assumed that it is possible to draw independent samples
under the null for each hypothesis in order to approximate its p-value.
Our aim is to use Monte Carlo samples to approximate the analytical test result (rejections and non-rejections),
obtained if all p-values were known, as accurately as possible.

This article proposed to use an idea based on \cite{Thompson1933} Sampling
to efficiently allocate samples to multiple hypotheses.
Our iterative \texttt{QuickMMCTest} algorithm is based on this principle
and adaptively allocates more samples to hypotheses whose decisions
are still unstable at the expense of allocating less samples to hypotheses
whose decision can easily be computed.
The algorithm works for a variety of common multiple
testing procedures for both a constant as well as a variable testing threshold.

\texttt{QuickMMCTest} has two main features:
First, it never computes any p-value estimates during its run, thus
avoiding to incur consistently non-rejecting all hypotheses as observed in other
methods published in the literature.
Second, its final decisions are based on empirical rejection probabilities
instead of p-value estimates.

\texttt{QuickMMCTest} was evaluated in a simulation study.
By comparing its performance to both a widely used naive sampling method
for a variety of commonly used multiple testing procedures
as well as to a variety of algorithms published in the literature,
we demonstrated that \texttt{QuickMMCTest} yields meaningful test results even at a low computational effort
and up to a multiple fold decrease in the number of switched classifications at a high effort.
For a low computational effort, \texttt{QuickMMCTest} yields a higher per-pair power
across all methods considered in this study and,
apart from the algorithm of \cite{GuoPedadda2008},
a lower false non-discovery proportion
when employed with various multiple testing procedures.

\section*{Supplementary Material}
The Supplementary Material compares the performance
of both \texttt{QuickMMCTest} variants using p-value estimates and empirical rejection probabilities.
Moreover,
it contains further simulation studies assessing
the dependence of \texttt{QuickMMCTest} on its parameters:
the total number of samples $K$ to be spent,
the number of updates $n_{\max}$ (and thus the number of samples $\Delta=K/n_{\max}$ spent in each iteration)
as well as the parameter $R$ controlling the accuracy with which weights are computed.
Moreover, the Supplementary Material
repeats the simulation studies
conducted in Section \ref{section_simulation_study}
for the variable testing threshold of \cite{PoundsCheng2006}.

\section*{Acknowledgements}
We would like to thank the two referees
for their constructive comments on the manuscript.
The second author was supported by the EPSRC.

\end{document}